\documentclass[a4paper,12pt,oneside]{article}

\usepackage{amsmath,amssymb,graphicx}
\usepackage{color,graphicx}

\usepackage{comment}

\def\lk{\left\{}

\def\z{\zeta}

\def\om{\omega}
\def\lp{\left(}
\def\rp{\right)}
\def\lbr{\left[}
\def\rbr{\right]}
\def\be{\begin{equation}}
\def\ee{\end{equation}}
\def\ai{\'{\i}}
\def\ba{\begin{array}}
\def\ea{\end{array}}
\def\bea{\begin{eqnarray}}
\def\eea{\end{eqnarray}}

\def\lb{\label}
\def\be{\begin{equation}}
\def\ee{\end{equation}}
\def\d{\delta}
\def\f{\phi}
\def\pd{\partial}

\usepackage{mathrsfs}
\usepackage{amsmath}
\usepackage{amssymb}
\usepackage{graphicx}
\usepackage{tabularx}
\usepackage[latin1]{inputenc}
\usepackage{enumerate}
\usepackage{graphicx,amssymb,amstext,amsmath}
\usepackage[bottom=3.5cm, textheight=22.5cm, textwidth=15cm]{geometry}
\usepackage{cancel}
\usepackage[para]{footmisc}

\begin{document}

\baselineskip0.25in
\title{Electrostatics and self-force in asymptotically flat cylindrical wormholes}

\author{E. Rub\ai n de Celis$^{1,2}$\footnote{e-mail: erdec@df.uba.ar}\, and C. Simeone$^{1,2}$\footnote{e-mail: csimeone@df.uba.ar}\\
{\footnotesize   $^1$ Departamento de F\ai sica, Facultad de Ciencias Exactas y Naturales,} 
\\
{\footnotesize Universidad de Buenos Aires, Ciudad Universitaria Pab. I, 1428,Buenos Aires, Argentina.}
\\
{\footnotesize $^2$ IFIBA--CONICET, Ciudad Universitaria Pab. I, 1428, Buenos Aires, Argentina.}}
\maketitle

\begin{abstract}

The problem of the electrostatics in conical wormholes is revisited, now improving the background geometries with asymptotical flatness. The electric self-force on a point charge placed at different regions in the spacetime of a conical thin-shell wormhole connecting flat outer submanifolds is obtained and compared with the results of previous works. The study is also carried out in terms of a previously introduced analogy in which the effect of the matter shells on the electric field is reproduced by non-gravitating layers of charge located on the boundary surfaces. Besides, a better insight on the physical effects of a non trivial topology is obtained by means of a further analysis of the electric fluxes across the wormhole throat and at both spatial infinities.
It is found that the throat is traversed by a non-arbitrary and finite topological flux, proportional to the charge of the source, which is characteristic of the asymptotically flat cylindrical wormholes regardless of the details of the throat
geometry.
\\

\end{abstract}

\section{Introduction}

Amongst other remarkable physical predictions, the relativistic theory of gravity introduces the possibility of a self-force on an  electric charge, as a result of the anisotropy of the Maxwell field induced by the non-Minkowskian character of the background geometry associated to a matter source. Early works showed this very interesting effect for rest charges in black hole backgrounds, both in a weak field approximation \cite{vil79} as in the context of the full field equations \cite{will80}\footnote{See \cite{Krtous:2019uqr}, and references therein, to find different approaches to the problem in black holes backgrounds.}.  The Maxwell equations for curved spacetime \cite{ll} (with $c=1$) are:
\begin{eqnarray}
\left(F^{\mu\nu}\sqrt{-g}\right)_{,\nu} & = & 4\pi j^\mu \sqrt{-g}\nonumber\\
F_{\mu\nu} & = & A_{\nu,\mu}-A_{\mu,\nu}\nonumber\\
j^\mu & = & \sum_a\frac{q_a}{\sqrt{-g}}\delta({\bf x}-{\bf x}_a)\frac{dx^\mu}{dx^0}\,,
\end{eqnarray}
where $\mu=(0,1,2,3)$, $g$ is the determinant of a given background metric defined by the tensor $g_{\mu\nu}$, $A_\mu$ is the four-potential, $j_\mu$ is the four-current and $q_a$ are the associated  charges. For only one rest charge $q$ located at the point ${\bf x}'$ we simply have
\begin{equation}\label{ff}
 \left(F^{0k}\sqrt{-g}\right)_{,k}=4\pi q\delta({\bf x}-{\bf x}') \, ,
\end{equation}
where $k=(1,2,3)$. These equations, together with the suitable boundary conditions corresponding to the particular background spacetime considered, completely determine the field of a point charge. Boundary conditions play a central role in the problem, as they explain the possibility of an anisotropic field around a rest point charge located in a region which is locally flat, and, consequently, the existence of an electric self-force in such an, at a first sight, trivial situation. A well known example of this is the self-force on a point charge in the conical --and thus locally flat-- spacetime of a gauge cosmic string \cite{vilenkin,vilgothis1,vilgothis2,vilgothis3} early calculated by Linet \cite{lin}. Electrostatic self-forces in wormhole \cite{visser} spacetimes were first studied in Refs. \cite{whs2,whs2.1,whs3}. Taking into account the interesting results therein, in previous works we developed a program aimed to distinguish between topologically different geometries which are equal in the region where the charge is placed.  We have then addressed the determination of the self-force in cylindrical thin-shell spacetimes \cite{prd12,epjc16} and also in spherically symmetric ones \cite{prd13} or different number of spacetime dimensions \cite{prd15}. Our program has also been extended to the case of scalar massless and massive fields \cite{epjc18,epjc19}.  

The particular class of cylindrically symmetric geometries as those associated to cosmic strings has received a renewed attention in the last decades \cite{strings1,strings2,strings3,strings4,strings5,strings6}, given, among other reasons, their possible role as secondary seeds for structure formation in the Universe. Bronnikov and other authors have pointed the convenience of asymptotical flatness in axially symmetric problems \cite{br13,br16,br19}; such feature can be achieved by suitably matching a geometry with a nontrivial far behavior to a flat one, thus allowing to turn geometries with unusual asymptotics into ones with a physically more appealing far behavior. Following this proposal, here we are interested in the problem of determining the electric field of a rest point charge in the background spacetime of a wormhole  connecting two gauge cosmic string geometries at each side of the throat, joined to two Minkowski geometries in their corresponding outer regions. Therefore, in the present work we will determine how this improvement in the kind of background geometries considered is reflected in the behavior of the electric self-force on a point charge, and how this compares with our previous results obtained for everywhere conical spacetimes \cite{prd12,epjc16}. We will also consider an interpretation at the light of the analogy introduced in \cite{epjc16}, where the effect of a cylindrical thin matter layer on the electric field is reproduced by replacing such shell by a suitable --non-gravitating-- charge density placed at the corresponding boundary.
 
Finally, a detailed analysis of the electric fluxes across the wormhole throat as well as at both spatial infinities will be performed; this will give a good picture of the interesting physical aspects introduced by the non trivial topology of the geometry. Unlike what was pointed out in previous illuminating works on electrostatics in spherically symmetric wormhole spacetimes \cite{whs3,whs2.1}, here we show that the presence of a test electric charge fixes a finite flux traversing the cylindrical wormhole throat. 

\section{The geometry}

Cylindrical wormholes of the thin-shell class, including those connecting gauge cosmic string submanifolds, i.e. conical geometries, have been extensively studied \cite{cilts1,cilts21,cilts22,cilts23,cilts24,cilts25,Mazharimousavi:2014gpa,Setare:2014eba,cilts26,Zaeem-ul-HaqBhatti:2017kaw,cilts27,Forghani:2018fks}. Now we consider an extension of the latter, that is an asymptotically flat thin-shell wormhole (FW) manifold  defined as 
\be
\mathcal{M}=\mathcal{M}_{e-}\cup\mathcal{M}_{i-}\cup\mathcal{M}_{i+}\cup\mathcal{M}_{e+},
\ee
where the inner submanifolds
\be
\mathcal{M}_{i\pm}=\{{\bf x}/a 
\leq r_\pm 
\leq b_{i
}
\}
\ee
have conical metrics, and the exterior ones
\be
\mathcal{M}_{e\pm}=\{{\bf x}/b
\leq r_\pm < \infty\}
\ee
are Minkowski geometries; the wormhole throat is located at $r_\pm = a$, while $b_{i}$ and $b$ are the radial coordinates at each side of the outer joining surfaces present in both, the {\it minus} and the {\it plus}, submanifolds of the spacetime. This spacetime is supported by thin layers of matter located at these three joining surfaces. 
The novelty in the present analysis clearly relies in the asymptotic flatness, so we are not going to mess with  the possibility of different geometries at each side of the wormhole throat; thus we have two identical submanifolds $\mathcal{M}_-=\mathcal{M}_{e-}\cup\mathcal{M}_{i-}$ and $\mathcal{M}_+=\mathcal{M}_{i+}\cup\mathcal{M}_{e+}$ with metrics \cite{vilgothis1,vilgothis2,vilgothis3}
\be
ds_{\pm}^2=-dt^2+dr^2+\rho^2(r)d\phi^2+dz^2,
\ee
\be
 \rho(r) = \left\{ \ba{ll}
          \, \om r \, , & \mbox{if $a\leq r \leq b_i $ \ \ \ ($\mathcal{M}_{i\pm}$)}\,, \\
 r\,, & \mbox{if $b \leq r < \infty$\ \ \  ($\mathcal{M}_{e\pm}$)}\,.
         \ea \right.
\ee
As usual, we have
$$-\infty<t<+\infty\,,  \quad0\leq\f\leq2\pi\,,\quad -\infty<z<+\infty \,.$$
In the case of the cosmic string geometries from which we start our mathematical construction, and  in units such that $G=1$, the parameter $\om$ would be given by $\om=1-8\mu$ where $\mu$ is the mass per unit length of a gauge cosmic string. Its associated stress-energy tensor has the form $T_\gamma^\nu=(-\rho,\,p_r\,,p_\phi\,,p_z)=-\mu\,\mathrm{diag}(1, 0,0,1)\,\delta(x)\,\delta(y)$, so that the pressure along the axis is negative, i.e. it is really a tension. Here we only assume $0<\om\leq 1$.

The continuity of the geometry at the throat is automatically satisfied, while across both outer joining surfaces it requires that we define the  coordinates at each side so that $\om b_i = b $. While the geometry must be continuous, the derivatives of the metric are not forced to do so. In general, the presence of a thin layer of matter will be associated to a jump in these derivatives. When put in covariant form, this relation between the geometry and the surface matter  has the form of the Lanczos equations \cite{daris1,daris2,daris3}
\be  \lb{le2}
8\pi S_{ij}  = [K] h_{ij} - [K_{ij} ],
\ee
where  $h_{ij}$ is the induced three--dimensional metric on the joining surface, $[K_{ij}] = K_{ij}^{(2)} - K_{ij}^{(1)}$ is the discontinuity of the extrinsic curvature tensor across this surface, $[K] = h^{ij} [K_{ij}]$ is its corresponding trace,  and $S_{ij}$ is the stress-energy tensor of the matter shell. In mixed components, the surface stress-energy tensor for a cylindrical shell has the form 
\be
S_i^j = \mathrm{diag}(-\sigma, p_\phi,p_z),
\ee
with $\sigma$ the surface energy density, and $p_\phi$ and $p_z$ the surface pressures. For a static configuration, at the wormhole throat we have 
\be 
\sigma=   -\frac{1}{4 \pi a},
\ee
\be 
p_\phi = 0
\ee
and
\be 
p_z = \frac{1}{4 \pi a}.
\ee
The surface energy density and pressure on each outer shell are given by 
\be 
\sigma=    \frac{\om-1}{8 \pi b} ,
\ee
\be 
p_\phi = 0
\ee
and
\be 
p_z = \frac{1-\om}{8 \pi b} .
\ee
Note that in all cases the energy density is negative, thus reflecting the presence of exotic matter. Such matter at the throat is necessary to have a wormhole geometry, while that at the outer joining surfaces is the price to be paid for asymptotical flatness. Another aspect to be mentioned is that the relation between the energy density and pressures on each shell is the same as for a gauge cosmic string; however, while for a string (of course with positive energy density) such relation implies a tension along the axis, for the shells in our construction the situation is reversed, and we effectively have a pressure in the direction parallel to the axis of symmetry.  

\section{Field of a point charge}

Starting from the background geometry introduced above, we now determine the electric field of a rest point charge, neglecting any backreaction of this field on the spacetime metric. In general, for a static point charge located at  $(r',\phi',z')$ in a conical spacetime with angle deficit $2\pi(1-\om)$, from (\ref{ff}) we can write down the following equation
\be
\nabla^2 A^0=-\frac{4 \pi q}{\om r}\,\d(r-r')\d(\phi-\phi')\d(z-z')
\ee
for the only non vanishing component  $A^0 \equiv V$  of the four-potential, where $\nabla^2\equiv g^{kl}\nabla_k\nabla_l$; the other components of the four-potential vanish. The explicit form of this equation is
\be\label{lap}   
\left[\frac{\partial^2}{\partial r^2} + \frac{1}{r} \frac{\partial}{\partial r} + \frac{1}{\om^2r^2} \frac{\partial^2}{\partial \phi^2} + \frac{\partial^2}{\partial z^2 } \right] V = -\frac{4 \pi q}{\om r}\,\d(r-r')\d(\phi-\phi')\d(z-z').
\ee
As is usual in analogous problems, we look for a solution of the form 
\be 
V = \frac{q}{ \om  } \frac{4}{\pi } \sum_{n=0}^{+\infty} \frac{\cos[n(\phi-\phi')]}{1+\d_{0,n}}  \int \limits_{0}^{+\infty}dk\,\cos[k(z-z')]\chi_{n}(k,r)  \, ,
\ee
where $F_z=\{ \cos[k(z-z')]\}$ and $F_{\phi}=\{ \cos[n(\phi-\phi')]\}$ are a complete set of orthogonal functions of the coordinates $z$ and $\phi$. By substituting this expression for $V$ in Eq. (\ref{lap}) we obtain the following equation for the radial functions $\chi_{n}(k,r)$:
\be 
\left\{ \frac{\partial}{\partial r} \left[ r \frac{\partial}{\partial r} \right] - r\left[ \lp \frac{n}{\om r} \rp^2 + k^2  \right]\right\} \chi_n(k,r) =  - \delta(r - r') \,.
\ee
The steps above will be applied to calculate the potential for a point charge in different cases. Then we shall identify the part of the field which is regular at the position of the charge, which is the contribution determining the electric self-force. 

We shall first study the case of a point charge between $a$ and $b_i$ because we are interested in the comparison with the example, already studied, of the field of  a charge in an everywhere conical thin-shell wormhole background. We shall assume that the charge is placed in the $\mathcal{M}_{i+}$ submanifold region. The potential $V_{FW}$  for a charge $q$ located at $(r',\f',z') \,\epsilon\,\mathcal{M}_{i+}$ is given by the solution of the following equations valid for the corresponding regions of the complete manifold:
\be\lb{mw}\ba{ll}
\nabla^2_{(\om)} V_{FW}^{i\pm} =
\left[\frac{\pd^2}{\pd r^2}+\frac{1}{r}\frac{\pd}{\pd r}+
\frac{1}{\om^2 r^2}\frac{\pd^2}{\pd \f^2}+\frac{\pd^2}{\pd z^2}\right] V_{FW}^{i\pm}
\qquad\qquad\qquad\qquad\qquad\qquad \\\\
\qquad\qquad\;\;\, =\left\{\ba{ll}
-\frac{4 \pi q}{\om}\,\d(r-r')\d(\phi-\phi')\d(z-z')& \mbox{for $\mathcal{M}_{i+}$} \, ,\\
0& \mbox{for $\mathcal{M}_{i-}$}\,.
\ea\right.
\ea
\ee

\be\lb{mw}\ba{ll}
\nabla^2 V_{FW}^{e\pm} =
\left[\frac{\pd^2}{\pd r^2}+\frac{1}{r}\frac{\pd}{\pd r}+
\frac{1}{ r^2}\frac{\pd^2}{\pd \f^2}+\frac{\pd^2}{\pd z^2}\right] V_{FW}^{e\pm}
\qquad\qquad\qquad\qquad\qquad\qquad \\\\
\qquad\qquad\;\;\, =\left\{\ba{ll}
0& \mbox{for $\mathcal{M}_{e+}$} \, ,\\
0& \mbox{for $\mathcal{M}_{e-}$}\,.
\ea\right.
\ea
\ee

The boundary conditions to be satisfied by the potential are:

\begin{itemize}

\item From the periodicity condition of the potential on the angle $\phi$:
\begin{enumerate}
\item $V_{FW}^{i\pm}(r,\phi=0,z)=V_{FW}^{i\pm}(r,\phi=2\pi,z)$,
\item $\frac{\pd}{\pd \phi}V_{FW}^{i\pm}(r,\phi=0,z)=\frac{\pd}{\pd \phi}V_{FW}^{i\pm}(r,\phi=2\pi,z)$.
\item $V_{FW}^{e\pm}(r,\phi=0,z)=V_{FW}^{e\pm}(r,\phi=2\pi,z)$,
\item $\frac{\pd}{\pd \phi}V_{FW}^{e\pm}(r,\phi=0,z)=\frac{\pd}{\pd \phi}V_{FW}^{e\pm}(r,\phi=2\pi,z)$
\end{enumerate}
\item From the globally valid continuity of the potential:
\begin{enumerate}
\item  $V_{FW}^{i+} \left(r\rightarrow r'^+,\phi,z\right)=V_{FW}^{i+} \left(r \rightarrow r'^-,\phi,z\right)$,
\item  $V_{FW}^{i+}(r\rightarrow a^+,\phi,z)=V_{FW}^{i-} \left(r \rightarrow a^+,\phi,z\right)$.
\item  $V_{FW}^{i\pm}(r\rightarrow b_i^-,\phi,z)=V_{FW}^{e\pm} \left(r \rightarrow b^+,\phi,z\right)$.
\end{enumerate}
\item The admissible asymptotic behavior implies:
\begin{enumerate}
\item $\lim \limits_{r \rightarrow \infty}V_{FW}^{e\pm}=0$.
\end{enumerate}
\item Requirements on the slope of the potential coming from the continuity of the associated electric field at the joining surfaces and the discontinuity of the field at the point charge source determine that:
\begin{enumerate}
\item $\frac{\pd}{\pd r}V_{FW}^{i+}(r\rightarrow a^+,\phi,z)=-\frac{\pd}{\pd r}V_{FW}^{i-}(r\rightarrow a^+,\phi,z),$
\item $\frac{\pd}{\pd r} V_{FW}^{i+}(r \rightarrow r'^+,\phi,z)- \frac{\pd}{\pd r}V_{FW}^{i+}(r \rightarrow r'^-,\phi,z) =
-\frac{4\pi q}{\om}\d(\phi-\phi')\d(z-z') \, ,$
\item $\frac{\pd}{\pd r}V_{FW}^{i\pm}(r\rightarrow b_i^-,\phi,z)=\frac{\pd}{\pd r}V_{FW}^{e\pm}(r\rightarrow b^+,\phi,z),$
\end{enumerate}
\end{itemize}

From these boundary conditions we can obtain the radial functions $\chi_{n}^{(i,e)\pm}(k,r)$ of the potential in each region. We are particularly interested in the potential for the region where the charge is placed; the radial functions for the charge in an inner region $\mathcal{M}_{i+}$ are
\be
\chi_{n}^{i+}(k,r) =  A_{n}(k) K_{\frac{n}{\om}}(k r) + B_{n}(k) I_{\frac{n}{\om}}(k r) + I_{\frac{n}{\om}}(k r_{<}) K_{\frac{n}{\om}}(k r_{>}) 
\ee
where $r_{><} = \{ r', r\}$, the coefficients are
\be
A_{n}(k) = - P_{n}(kr') \frac{\left( P_{n}(k a) I_{\frac{n}{\om}}(k a)\right)'}{2 P_{n}(k a) P'_{n}(k a)}
\ee
\be
B_{n}(kr) = \left( A_{n}(k) + I_{\frac{n}{\om}}(k r') \right) G_n(k) 
\ee
with the auxiliary definitions 
\be \label{P_n(kr)}
P_{n}(kr) \equiv I_{\frac{n}{\om}}(k r) G_n(k) + K_{\frac{n}{\om}}(k r) 
\ee
\be
G_n(k) \equiv - \frac{K'_{\frac{n}{\om}}(k b_{i}) K_{n}(k b) - K_{\frac{n}{\om}}(k b_{i}) K'_{n}(k b)}{I'_{\frac{n}{\om}}(k b_{i}) K_{n}(k b) - I_{\frac{n}{\om}}(k b_{i}) K'_{n}(k b)};
\ee
where the primes over functions imply a derivative with respect to the corresponding radial argument $a$, $b_{i}$ or $b$.
\\

Now we turn to the case of  a point charge in the outer region $r > b$; we evaluate $V_{FW}$ for a charge $q$ at $(r',\f',z') \,\epsilon\,\mathcal{M}_{e+}$. 
This would allow the comparison, for instance, with the field of a charge in a flat wormhole spacetime (that is, a wormhole joining two Minkowski geometries). The potential is determined by the solution of the following equations corresponding to the four submanifolds which constitute the complete manifold $\mathcal{M}$:
\be\lb{mw}\ba{ll}
\nabla^2_{(\om)} V_{FW}^{i\pm} =
\left[\frac{\pd^2}{\pd r^2}+\frac{1}{r}\frac{\pd}{\pd r}+
\frac{1}{\om^2r^2}\frac{\pd^2}{\pd \f^2}+\frac{\pd^2}{\pd z^2}\right] V_{FW}^{i\pm}
\qquad\qquad\qquad\qquad\qquad\qquad \\\\
\qquad\qquad\;\;\, =\left\{\ba{ll}
0& \mbox{for $\mathcal{M}_{i+}$} \, ,\\
0& \mbox{for $\mathcal{M}_{i-}$}\,.
\ea\right.
\ea
\ee

\be\lb{mw}\ba{ll}
\nabla^2_{} V_{FW}^{e\pm} =
\left[\frac{\pd^2}{\pd r^2}+\frac{1}{r}\frac{\pd}{\pd r}+
\frac{1}{ r^2}\frac{\pd^2}{\pd \f^2}+\frac{\pd^2}{\pd z^2}\right] V_{FW}^{e\pm}
\qquad\qquad\qquad\qquad\qquad\qquad \\\\
\qquad\qquad\;\;\, =\left\{\ba{ll}
-\frac{4 \pi q}{r}\,\d(r-r')\d(\phi-\phi')\d(z-z')&\mbox{for $\mathcal{M}_{e+}$} \, ,\\
0& \mbox{for $\mathcal{M}_{e-}$}\,.
\ea\right.
\ea
\ee

The boundary conditions to be satisfied by the potential are:

\begin{itemize}

\item From the periodicity of the potential on the angular coordinate $\phi$:
\begin{enumerate}
\item $V_{FW}^{i\pm}(r,\phi=0,z)=V_{FW}^{i\pm}(r,\phi=2\pi,z)$,
\item $\frac{\pd}{\pd \phi}V_{FW}^{i\pm}(r,\phi=0,z)=\frac{\pd}{\pd \phi}V_{FW}^{i\pm}(r,\phi=2\pi,z)$.
\item $V_{FW}^{e\pm}(r,\phi=0,z)=V_{FW}^{e\pm}(r,\phi=2\pi,z)$,
\item $\frac{\pd}{\pd \phi}V_{FW}^{e\pm}(r,\phi=0,z)=\frac{\pd}{\pd \phi}V_{FW}^{e\pm}(r,\phi=2\pi,z)$.
\end{enumerate}
\item From the continuity of the potential, valid everywhere:
\begin{enumerate}
\item $V_{FW}^{e+} \left(r\rightarrow r'^+,\phi,z\right)=V_{FW}^{e+} \left(r \rightarrow r'^-,\phi,z\right)$,
\item $V_{FW}^{i+}(r\rightarrow a^+,\phi,z)=V_{FW}^{i-} \left(r \rightarrow a^+,\phi,z\right)$.
\item $V_{FW}^{i\pm}(r\rightarrow b_i^-,\phi,z)=V_{FW}^{e\pm} \left(r \rightarrow b^+,\phi,z\right)$
 \end{enumerate}
\item The requirement of a good asymptotic behavior implies
\begin{enumerate}
\item $\lim \limits_{r \rightarrow \infty}V_{FW}^{e\pm}=0$.
\end{enumerate}
\item Conditions on the slope of the potential dictated by a discontinuity of the associated electric field at the point charge source and continuity at avery other point, in particular at the joining surfaces, imply the equalities:
\begin{enumerate}
\item $\frac{\pd}{\pd r}V_{FW}^{i+}(r\rightarrow a^+,\phi,z)=-\frac{\pd}{\pd r}V_{FW}^{i-}(r\rightarrow a^+,\phi,z),$
\item  $\frac{\pd}{\pd r} V_{FW}^{e+}(r \rightarrow r'^+,\phi,z)- \frac{\pd}{\pd r}V_{FW}^{e+}(r \rightarrow r'^-,\phi,z) =
-\frac{4\pi q}{r}\d(\phi-\phi')\d(z-z') \, ,$
\item  $\frac{\pd}{\pd r}V_{FW}^{i\pm}(r\rightarrow b_i^-,\phi,z)=\frac{\pd}{\pd r}V_{FW}^{e\pm}(r\rightarrow b,^+\phi,z),$
\end{enumerate}
\end{itemize}

Again, we can obtain the radial functions $\chi_{n}^{(i,e)\pm}(k,r)$ of the potential in each region from these boundary conditions. In particular, the radial functions for the charge in the outer region $\mathcal{M}_{e+}$ are 
\be \label{chi_e}
\chi_{n}^{e+}(k,r) = C_{n}(k) K_{n}(k r) + I_{n}(k r_{<}) K_{n}(k r_{>}) 
\ee
where $r_{><} = \{ r', r\}$, the coefficients read
\be
C_{n}(k) = - K_{n}(kr') \frac{I'_{n}(k b) W_{n}(k b_i) - I_{n}(k b) W'_{n}(k b_i)}{K'_{n}(k b) W_{n}(k b_i) - K_{n}(k b) W'_{n}(k b_i) }, 
\ee
and the auxiliary functions $W_n$ are given by
\be
W_{n}(k r) \equiv K_{\frac{n}{\om}}(k r) (P_n(k a) I_{\frac{n}{\om}}(k a))' -  I_{\frac{n}{\om}}(k r) (P_n(k a) K_{\frac{n}{\om}}(k a))'
\ee
with $P_n(k a)$ defined as in (\ref{P_n(kr)}), and  the prime over functions implying derivative with respect to the corresponding radial argument $a$, $b_{i}$ or $b$.
\\

\section{Evaluation of the self-force}

For the evaluation of the self-force we need the potencial at the position of the charge where it admits to be written as the superposition
\be \lb{superp}
V_{FW} = V_{bulk} + V_{shells} \,,
\ee
with $V_{bulk}$ containing the inhomogeneity at $r=r'$ and $V_{shells}$ containing the information corresponding to the distortions coming from the boundary conditions at the shells and from the topology of the wormhole\footnote{An interpretation in terms of an analogous problem where the effect of the matter shell is reproduced by a non-gravitating layer of charge on the boundary surface can be seen in Ref. \cite{epjc16}. A central point in this analogy is that the charge density concentrated on the region of the surface which is near to the charge location has equal (opposite) sign to that of the point charge according to the normal (exotic) nature of the matter shell, producing a repulsive (attractive) effect. See also the discussion below.}. The divergent term $V_{bulk}$ is locally equivalent to the potential in the spacetime of a cosmic string with parameter $\om \in (0;1]$, i.e., $V_{bulk} = V_{CS}$. The renormalization of our potential function at the position $r=r'$ is, therefore, 
\be\lb{ren}
V^{ren}_{FW}(r') = V^{ren}_{CS}|_{r'} + V_{shells}|_{r'}\,,
\ee
where \cite{fulling} 
\be\lb{pot_cs}
V_{CS} =  \frac{q}{\pi \sqrt{2 r r'}} \int \limits_{u}^{+\infty}\frac{ \sinh(\zeta/\om) \; \om^{-1} \quad d\z}{\left[ \cosh(\z/\om) - \cos \frac{\phi - \phi'}{\om} \right] \lp \cosh \z - \cosh u \rp^{1/2}}   
\ee
with
\be
\cosh u = \frac{r^2 + r'^2 + (z-z')^2}{2 r r'} \,, \quad u  \geqslant 0\,.
\ee
The geometry is locally flat in a neighborhood of the charge, which means that the singular part $V_S$ to be subtracted is identical to the potencial in a Minkowski spacetime; hence $V_S$ can be written by putting $\om = 1$ in the previous expression: 
\be
V_S = \frac{q}{\pi} \frac{1}{\sqrt{2 r r'}} \int \limits_{u}^{+\infty}  \frac{\sinh\z \quad {d\z}}{\left[ \cosh\z - \cos (\phi - \phi') \right] \lp \cosh \z - \cosh u \rp^{1/2}}
\,.
\ee
Evaluating at $z=z'$ and $\phi=\phi'$ to take the coincidence limit using radial geodesics we obtain
\bea \lb{Lin}
V^{ren}_{CS} &=&  \lim_{r \to r'} \lp V_{CS}\lp r,r' \rp - V_S({r},{r'})  \rp \nonumber \\ 
&=& \frac{q}{2\pi} \frac{L_{\om}}{r'}\,, \quad \mbox{with:} \qquad\qquad\qquad
\eea
\be \lb{coefL}
L_{\om}=\int \limits_{0}^{+\infty} \left[ \frac{\sinh(\zeta/\om)}{\om \left[\cosh(\zeta/\om)-1\right]}-\frac{\sinh{\zeta}}{\cosh{\zeta}-1}\right]\frac{d\zeta}
{\sinh(\zeta/2)} \,.
\ee
Finally, the regularized potential of the inhomogeneous term in a conical geometry is 
\be \lb{regCS}
V^{reg}_{CS} (r,r') = \frac{q}{4 \pi} \frac{L_{\om}}{r'} \lp 2 - \ln{\frac{r}{r'}} \rp \,,
\ee
from where one obtains $V^{ren}_{CS} = V^{reg}_{CS} (r,r') |_{r'} $. 
The regularized potential at the position of the charge at our wormhole geometry is $V^{reg}_{FW} =  V^{reg}_{CS} + V_{shells}$.
A term like (\ref{regCS}) will be present if the charge is placed in an angle deficit submanilod, i.e., in the interior regions of the wormholes where $\om < 1$.
In the flat exteriors, without angle deficit, we have $V^{reg}_{CS} = 0$ (which can be checked from the coefficient (\ref{coefL})). Consequently, we obtain
\be
V^{reg}_{FW} = \lk \ba{ll} 
   q
    \frac{L_{\om}}{4\pi} \frac{1}{r'} \lp 2 - \ln{\frac{r}{r'}} \rp + 
     \frac{q}{\om} \sum\limits_{n=0}^{\infty}  \frac{4}{\pi \lp 1+\delta_{n,0}\rp}  \int \limits_{0}^{+\infty} dk \, \lbr  K_{\frac{n}{\om}} (kr) A_n(k) +I_{\frac{n}{\om}} (kr) B_n(k)  \rbr
\,, 
\mbox{in}\; \mathcal{M}_{i} \\
 q \sum\limits_{n=0}^{\infty}  \frac{4}{\pi \lp 1+\delta_{n,0} \rp} \int \limits_{0}^{+\infty}dk \, K_{n} (k r) \, C_n(k)  \,, 
 \mbox{in}\;  \mathcal{M}_{e}
\ea \right.
\ee
for the interior and exterior regions, respectively. From the regularized potential we compute the self-force $\mathrm{f}$ on the charge $q$ as measured by a static observer in the wormhole geometry, $\mathrm{f} \equiv f^{\hat{r}} = - q \, \frac{\partial}{\partial_r} V_{FW}^{reg} |_{r=r'}$, this yields
\be \label{force}
\mathrm{f} = \lk \ba{ll} 
   q^2 \lbr
 \frac{L_{\om}}{4 \pi}   \frac{1}{r'^2} 
  -  \frac{1}{\om} \sum\limits_{n=0}^{\infty}  \frac{4}{\pi \lp 1+\delta_{n,0}\rp}  \int \limits_{0}^{+\infty} dk \, \lp K'_{\frac{n}{\om}} (kr')A_n(k)  + I'_{\frac{n}{\om}} (kr') B_n(k) \rp
  \rbr  \,,
  \mbox{in}\; \mathcal{M}_{i}  \\
-  \, q^2 \sum\limits_{n=0}^{\infty}  \frac{4}{\pi \lp 1+\delta_{n,0} \rp} \int \limits_{0}^{+\infty}dk \,   K'_{n} (k r')  \, C_n(k) \,,
\mbox{in}\;  \mathcal{M}_{e}
\ea \right. 
\ee
In an inner region $\mathcal{M}_{i}$ the force is described essentially by two parts; a positive term $\sim r'^{-2}$ which decreases with increasing $\om$, plus the contribution of the infinite series describing the deflection of the field generated by the boundary conditions. 
In the flat Minkowski exteriors $\mathcal{M}_{e}$ the force is produced exclusively by the effects involved at the boundaries. To represent the self-force we will plot it using the global dimensionless coordinate $x \in (-\infty; +\infty)$, defined as follows
\be 
r' = \lk
\ba{ll}
b \,x  \,, \; \mbox{if $+\infty > x \geqslant 1$}   \; \mbox{, in $\mathcal{M}_{e+}$}\\
 x \, \Delta+ a \,,\; \mbox{if $1 \geqslant x \geqslant 0$}\; \mbox{, in $\mathcal{M}_{i+}$} \\
 - x \, \Delta + a \,,\; \mbox{if $0 \geqslant x \geqslant -1$}\; \mbox{, in $\mathcal{M}_{i-}$} \\
- b \, x  \,, \; \mbox{if $-1 \geqslant x > -\infty$}   \; \mbox{, in $\mathcal{M}_{e-}$}
\ea 
\right.
\ee 
The parameter $\Delta \equiv b_i - a = b/\om -a $ is the size of each interior region or, equivalently, $2 \Delta$ is the size of the conical throat which joins the two Minkowski exteriors.  The position $x=0$ is the center of the throat of radius $a$, and the exterior shells are located to $x=\pm 1$.

\begin{figure} [h!]
\centering
  \begin{minipage}{0.33\textwidth}
    \centering
    \includegraphics[width=0.95\textwidth]{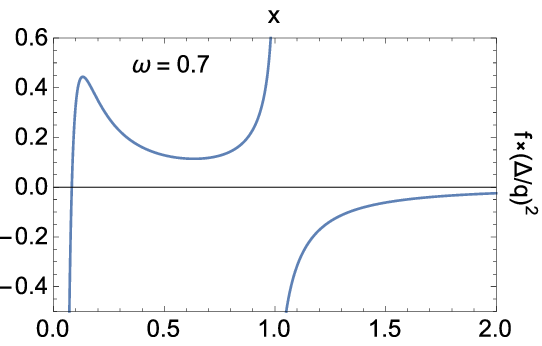} \\
    {\footnotesize (a) $\om=0.7$, $\Delta = 1000$, $a=1$.} 
  \label{fant}
  \end{minipage}%
       \begin{minipage}{0.33\textwidth}
    \centering
    \includegraphics[width=0.95\textwidth]{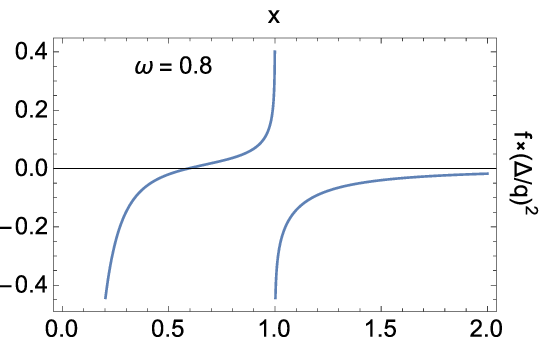}\\
    {\footnotesize (b) $\om=0.8$, $\Delta =1000$, $a=1$.} 
  \end{minipage}
  \begin{minipage}{0.33\textwidth}
    \centering
    \includegraphics[width=0.95\textwidth]{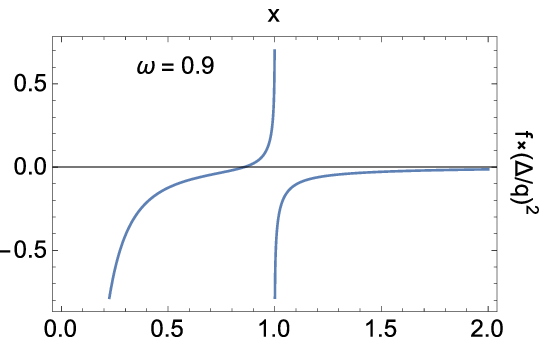}\\
    {\footnotesize (c) $\om=0.9$, $\Delta =1000$, $a=1$.} 
  \end{minipage}
  \caption{Dimensionless self-force $\mathrm{f} \times  (\Delta/q)^2$ in terms of the dimensionless coordinate $x$, for different values of the angle deficit parameter $\om$. The throat is at $x=0$ and the exterior shell is at $x=1$.}
  \label{vs_omega}
\end{figure}

In the following figures we plot $\mathrm{f} \times (\Delta/q)^2$ (dimensionless self-force), against the dimensionless coordinate $x$, for different values of the angle deficit parameter $\om$ of the conical interior, the size $\Delta$ and the radius $a$ of the throat.
Since the wormhole is symmetric across the throat, the force is represented for a charge placed in the {\it plus} region $\mathcal{M}_{+} =\mathcal{M}_{i+} \cup \mathcal{M}_{e+}$, only.
Figure \ref{vs_omega} shows the dependance on the deficit angle with the comparison of three cases with different $\omega$ and fixed $\Delta$ and $a$. As it could be expected from the first term in (\ref{force}), for a given size and radius of the throat, the repulsive self-force effect due to the conical geometry decreases for increasing $\om$.
\\
\begin{figure} [h!]
\centering
  \begin{minipage}{0.33\textwidth}
    \centering
    \includegraphics[width=0.95\textwidth]{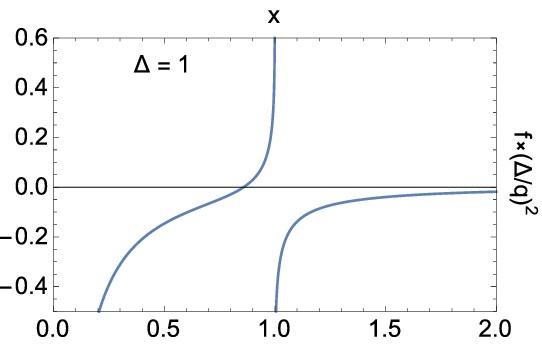}\\
    {\footnotesize (a) $\om=0.7$, $\Delta =1$, $a=1$.} 
  \label{fant}
  \end{minipage}%
  \begin{minipage}{0.33\textwidth}
    \centering
    \includegraphics[width=0.95\textwidth]{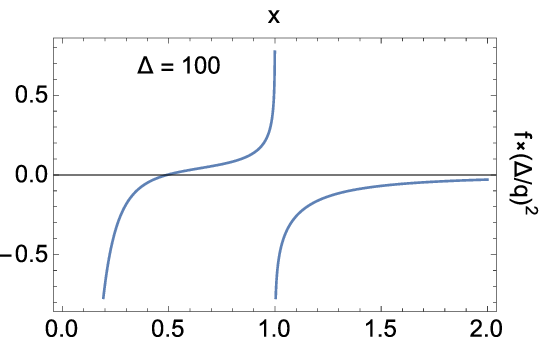}\\
    {\footnotesize (b) $\om=0.7$, $\Delta =100$, $a=1$.} 
  \end{minipage}
     \begin{minipage}{0.33\textwidth}
    \centering
    \includegraphics[width=0.95\textwidth]{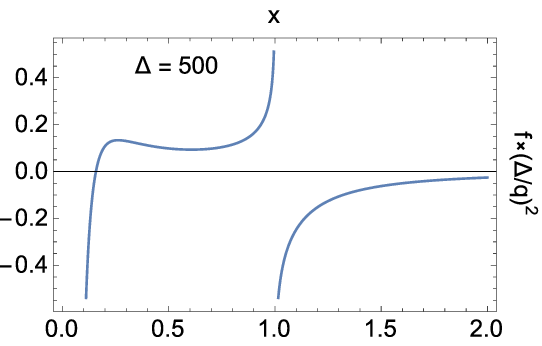}\\
    {\footnotesize (c) $\om=0.7$, $\Delta =500$, $a=1$.} 
  \end{minipage}
  \caption{Dimensionless self-force $\mathrm{f} \times (\Delta/q)^2$ in terms of the dimensionless coordinate $x$, for different sizes $\Delta$ of the conical region. The throat is at $x=0$ and the exterior shell is at $x=1$.}
  \label{vs_Delta}
\end{figure}
\begin{figure} [h!]
\centering
  \begin{minipage}{0.33\textwidth}
    \centering
    \includegraphics[width=0.95\textwidth]{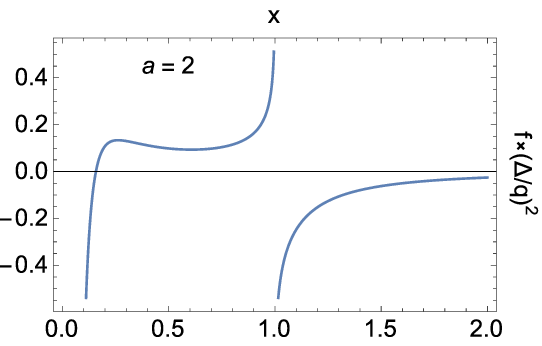}\\
    {\footnotesize (a) $\om=0.7$, $\Delta =1000$, $a=2$.} 
  \label{fant}
  \end{minipage}%
  \begin{minipage}{0.33\textwidth}
    \centering
    \includegraphics[width=0.95\textwidth]{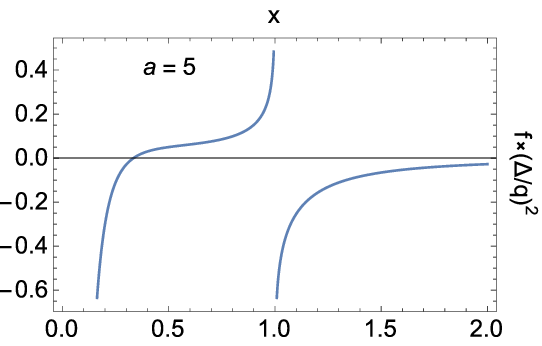}\\
    {\footnotesize (b) $\om=0.7$, $\Delta =1000$, $a=5$.} 
  \end{minipage}
     \begin{minipage}{0.33\textwidth}
    \centering
    \includegraphics[width=0.95\textwidth]{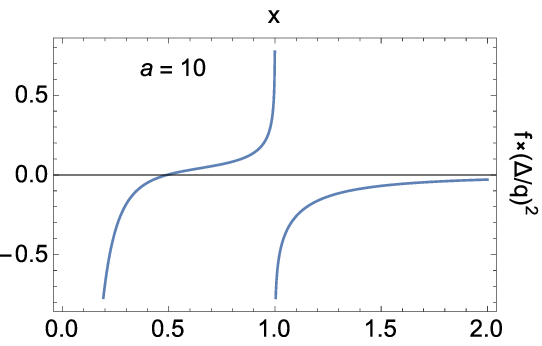}\\
    {\footnotesize (c) $\om=0.7$, $\Delta =1000$, $a=10$.} 
  \end{minipage}
  \caption{Dimensionless self-force $\mathrm{f} \times (\Delta/q)^2$ in terms of the dimensionless coordinate $x$, for increasing values of the throat radius $a$. The throat is at $x=0$ and the exterior shell is at $x=1$.}
  \label{vs_a}
\end{figure}

In figure \ref{vs_Delta} we show the dependance of the force on the size $\Delta$, while in figure \ref{vs_a} its dependance on the radius $a$. The analysis of the results shown in the figures reveals some points to be remarked:
\begin{enumerate}

\item Any shell with negative energy (exotic matter associated to a positive spatial curvature on a surface joining two submanifolds) has an attractive effect on the electric point charge: this happens near the shell at the wormhole throat as well as at both sides of the exotic outer shell required to achieve asymptotic flatness. Of course, for a charge in the conical region, if the position is near the throat this attraction points towards the axis, while for a charge located near the outer shell this attraction points outwards. The attractive effect of these exotic shells is studied in more detail below.

\item If the size $\Delta$ of the conical region is large enough so that the effect of the shells in the region between them (that is, between the throat and the outer shell) becomes comparatively weak, the dominant contribution to the electric self-force is a repulsion from the central axis associated to the angle deficit of the conical geometry. On the other hand, as this repulsive effect decays as $\sim 1/r'^{2}$ ($r'>a$), for larger values of the throat radius $a$ it eventually becomes negligible (see figures 2 and 3 with fixed $\om$).

\item In the outer flat regions, where no deficit angle exists, the only effect is the attractive self-force towards the axis; as pointed above, this force is a characteristic of the exotic matter shells which have positive spatial curvature on the joining surface.
\end{enumerate}

To understand the attractive contribution to the force produced by these shells of negative energy density we can apply the interpretation of the equivalent electric problem developed in \cite{epjc16}. 
This interpretation provides an alternative picture in which the actual deflection of the field lines generated at the shell's curvature jump is replicated with an analogous image method electric problem in which surface charge densities are placed at the boundaries between two regions. 
The analysis will show that, in addition to the deflection of the field by the thin-shells, the leading asymptotic attractive force towards the throat involves a drain hole effect which is characteristic of the wormhole topology.

Let us start considering the problem in an outer region $\mathcal{M}_{e+}$. If the charge is placed in $\mathcal{M}_{e+}$ the potential is the superposition of an inhomogeneous term produced by the particle plus an homogeneous part called $V_{shell}$, as pointed out earlier in equation (\ref{superp}).
The equivalent interpretation describes the problem of the point charge in the exterior Minkowski region of the wormhole using an equivalent electric potential $\varphi_e$ in a Minkowski spacetime in which the thin-shell contour at the surface $r=b$ is replaced by a non-gravitating surface charge distribution $\Sigma^b$. 
This charge density must produce a potential $\varphi_{b} \equiv V_{shell}$ in the region $b < r < \infty$. Defining $\varphi_{q}$ to be the inhomogeneous point charge electric potential in Minkowski space, the total field given by $\varphi_e = \varphi_{q} + \varphi_{b}$ is equivalent by construction to the original one in the outer region of the wormhole.
Following the procedure described in {\cite{epjc16}} (see equation (49) and explanation therein), the potential of the surface charge density is explicitly
\be
\varphi_{b} = q \int_{0}^{+\infty} dk\,\cos{[k(z-z')]} \sum_{n=0}^{\infty} 	\cos{[n(\phi-\phi')]} \frac{4 \, C_n(k)}{(1+\delta_{n,0})\pi\, I_n(kb)}  I_n(k r_{<}) K_n(k r_{>})
\,,
\ee
where $r_{><}=\{r,b\}$, and coincides with $V_{shell}$ in $b<r<\infty$, as given by the first term in (\ref{chi_e}).
Then, the associated surface charge density at $r=b$ is
\be \label{charge_density}
\Sigma^b(\phi,z) = \sum_{n=0}^{\infty} \cos{[n(\phi-\phi')]} \int_0^{\infty} dk \cos{[k(z-z')]} \Sigma_n^b(k) \,,
\ee
with the Fourier coefficients 
\be
\Sigma_n^b(k) = \frac{q}{4 \pi b}  \, \frac{4 \, C_n(k)}{(1+\delta_{n,0})\pi\, I_n(kb)} \,.
\ee
Like any induced surface charge, $\Sigma^b(\phi, z)$ depends on the position $(r',\phi', z')$ of the source charge $q$ and is well defined for $b < r' < \infty$ ($r'$ is implicit in the coefficient function $C_n(k)$). In the limit $r' \to b$ we have
\be
\lim_{r'\to b} \frac{\Sigma^b(\phi', z')}{q/\Delta^2} \,  = - \infty
\ee
as is expected from the numerical analysis in terms of the self-force, i.e. a peak of infinite opposite charge density is concentrated at $(\phi', z')$, correspondingly with the leading attraction to the exotic matter shell in the original problem.
An interesting calculation which can be performed using the equivalent electric problem is the net charge $Q_b$ of the effective surface distribution $\Sigma^b(\phi, z)$, this is
\begin{flalign}
Q_b = \int_{-\infty}^{+\infty} dz \int_0^{2\pi} b\, d\phi = 2 \pi^2  b\,
 \lim_{k\to0} \, \Sigma^b_0(k) =  - \frac{q}{2} \,.
  \end{flalign}
The above result points out that besides the particle's charge $q$, there is an effective net charge $Q_b=-q/2$ as seen from the exterior. 
This net charge of opposite sign is responsible for the leading asymptotic attraction of the particle towards the central axes.
Moreover, applying Gauss law for the net image charge there is an associated electric field flux equal to $4\pi Q_b$. 
Interpreted in terms the original problem, this flux is a drain effect towards the wormhole's inner region which is independent of the position of the particle with charge $q$.


Analogously to what was previously developed for the exterior, the equivalent problem can be posed in an interior region $\mathcal{M}_{i+}$.
Following the same procedure, we will consider two surface charge densities, one associated to each one of the boundaries of $\mathcal{M}_{i+}$.
The equivalent electric potential $\varphi_i = \varphi_q + \varphi_a + \varphi_{b_i}$ in the region $a<r<b_i$ of a conical geometry with parameter $\om$, is the superposition of: $\varphi_q$, the potential of the charge $q$ in the conical space, and $\varphi_a$ and $\varphi_{b_i}$ which are produced by charge densities at the surfaces $r=a$ and $r=b_i$, respectively.
The charge density $\Sigma_n^{a}(k)$ at $r=a$, and $\Sigma_n^{b_i}(k)$ at $r=b_i$ can be written as in (\ref{charge_density}), with the Fourier coefficients given by
\be
\Sigma_n^{a}(k) = \frac{q}{4 \pi a \om}  \, \frac{4 \, A_n(k)}{(1+\delta_{n,0})\pi\, I_{\frac{n}{\om}}(ka)} \,,
\ee
\be
\Sigma_n^{b_i}(k) = \frac{q}{4 \pi b_i \om}  \, \frac{4 \, B_n(k)}{(1+\delta_{n,0})\pi\, K_{\frac{n}{\om}}(k b_i)} \,,
\ee
respectively. We can check, for each, that 
\be
\lim_{r' \to \{a,b_i\}} 
\frac{\Sigma^{\{a,b_i\}}(\phi', z')}{q/\Delta^2} = - \infty
\ee
both negative, as it might be expected from the analyzed numerical results of the self-force. 
The net charges of these effective charge distributions are 
\begin{flalign}
Q_{a} = \int_{-\infty}^{+\infty} dz \int_0^{2\pi} a \om \, d\phi \, \Sigma^{a}(\phi, z) = 2 \pi^2  a\, \lim_{k\to 0} \, \Sigma^{a}_0(k)
 =  - \frac{q}{2} \,, 
  \end{flalign}
  \begin{flalign}
Q_{b_i} = \int_{-\infty}^{+\infty} dz \int_0^{2\pi} b_i\om \, d\phi \, \Sigma^{b_i}(\phi, z)	= 2 \pi^2  b_i\,
 \lim_{k\to 0} \, \Sigma^{b_i}_0(k) =  - \frac{q}{2} (1-\om) \,.
  \end{flalign}
We notice that across any surface enclosing the throat there is an electric flux $4\pi \,Q_a$ produced by the charge $Q_a =  -q/2$.
Interpreted in terms of the original problem, this flux enters across the throat, in the direction from $\mathcal{M}_-$ to $\mathcal{M}_+$,
acting as a drain hole effect for the particle with charge $q$ in the {\it plus} submanifold. \\
 
Independently of the position of the test charged particle, placed either in $\mathcal{M}_{e+}$ or $\mathcal{M}_{i+}$, we noticed an electric flux entering across the wormhole throat. The latter observation can be verified by calculating the electric flux through the throat at $r=a$; $\mathcal{F}_{wh} = \iint F_{\mu \nu} \, dX^{\mu \nu} $, using the potential $A_0=-V_{FW}$ for an arbitrary position of the particle in $\mathcal{M}_{+}$, and using the surface normal vector ${\bf r_-}$  pointing from {\it plus} to {\it minus} submanifold, this is
\be \label{F_wh}
\mathcal{F}_{wh}
= \iint_{throat} - {\bf \nabla} V_{FW} \, {\bf dS_-}
= 4\pi \, \frac{q}{2} \,.
\ee
The above is the characteristic drain hole effect produced by the presence of a test point charge $q$ in our asymptotically flat cylindrical wormholes. 
Regarding the imprint at infinity, we can check the electric field flux $\mathcal{F}_{\infty+}$ at the asymptotic boundary $\partial \mathcal{M}_{+}^{\infty}$ of the {\it plus} submanifold -where the source $q$ is placed in the  wormhole geometry- pointing in ${\bf r_+}$  direction
\be \label{F_+}
\mathcal{F}_{\infty+} = \iint_{\partial \mathcal{M}_{+}^{\infty}} - {\bf \nabla} V_{FW} \, {\bf dS_+} = 4\pi q - F_{wh}
= F_{wh}
=4\pi \, \frac{q}{2} \,,
\ee
which is in accordance with the previous calculations using the equivalent problem for the outer region, where we found $q+Q_b = q/2$. The corresponding imprint at the asymptotic infinity $\partial \mathcal{M}_{-}^{\infty}$ of the {\it minus} submanifold is checked to be $\mathcal{F}_{\infty-} = \mathcal{F}_{wh} =  4\pi \, q/2$, confirming the continuity of the electric field drain hole effect.
An important observation is that, independent of the position of the test charge $q$ in the wormhole spacetime, the total asymptotic flux is $\mathcal{F}_{\infty} = \mathcal{F}_{\infty-} +\mathcal{F}_{\infty+}  = 4 \pi q$, as it should be according to the total charge in the background geometry.


\section{Discussion}
 
When we compare the results with those previously obtained  we can first note that in the exterior problem (point charge in the Minkowski outer region beyond $r=b$), we obtain a self-force towards the axis which  qualitatively always reproduces the results shown in Fig. 4 of our previous work \cite{prd12}. We can say that the inclusion of a two-cones transition region between the two Minkowski sides of the wormhole does not introduce a central difference in the behavior of the electric self-force. For the interior problem (point charge in the conical region between $r=a$ and $r=b$), instead, the results are more interesting: the repulsion decaying with the radius found for large values of the radial coordinate of the charge in \cite{prd12} can now be found only for large values of $\Delta$, that is for the outer shell far from the wormhole throat, and for sensible amounts of angle deficit: see Figs. 1(a), 2(c) and 3(a) here, and Fig. 2 in \cite{prd12}. In general, the most interesting new aspect due to the presence of an outer shell is that for a charge in the conical region the self-force has a behavior which --differing from what happens without the outer shell-- can be somewhat described in terms of three sections: Near each shell the dominant contribution of the force corresponds to the attraction associated to their exotic nature, which in turn can be easily understood in terms of the analogous problem defined in \cite{epjc16}, where each exotic matter shell yields the same effect of a certain surface charge  of sign opposite to that of the point charge; hence for a charge near the throat we have a force towards the axis, while for a charge near the outer shell we have a force pointing outwards. For a charge in an intermediate zone, always within the conical region, the repulsive behavior of the self-force of the form $1/r^2$ due to the angle deficit of the cosmic string character of the geometry can be noted, though it can be clearly isolated only if the region is wide enough. In conclusion, global differences between geometries which can locally be equal, and consequently the existence of different regions beyond shells placed at the joining surfaces, can be revealed by observing the behavior of the electric self-force on a test charge. 
\\

In addition, we have also studied in detail the fluxes across the wormhole throat and at both spatial infinities, which provide a further understanding of the peculiarities introduced in the electrostatic problem by the non trivial topology of the background.
We found that the throat acts as a drain hole pierced by an electric flux $\mathcal{F}_{wh} = 4\pi \, q/2$ which points from the source's submanifold to the sourceless one; this is the direction from $\mathcal{M}_{+}$ to $\mathcal{M}_{-}$ in the calculations of the previous section in which we placed $q$ at the {\it plus} side.
With the latter configuration we obtained that the electric field flux pointing to the asymptotic {\it plus} region is $\mathcal{F}_{\infty+} = 4\pi \, q - \mathcal{F}_{wh} = 4\pi \, q/2$, and the flux pointing to the asymptotic {\it minus} submanifold is
$\mathcal{F}_{\infty-} = \mathcal{F}_{wh} = 4\pi \, q/2$.
This means that with the particle placed at $\mathcal{M}_{+}$, the net charge as seen from {\it plus} infinity is not $q$, it is $q/2$.
It is important to observe that the total asymptotic flux is $\mathcal{F}_{\infty} \equiv \mathcal{F}_{\infty-} + \mathcal{F}_{\infty+} = 4\pi q$, as it should be according to the total electric charge in the background wormhole geometry.

We say that the quantity $\mathcal{F}_{wh}$ is a topological flux because it is exclusively related to the particular non-trivial topology of the asymptotically flat cylindrical wormhole regardless of the details of the throat i.e.; it is independent of the radius $a$, the size $\Delta$, and the parameter $\om$ of the inner region.
We remark that it is not an arbitrary flux: it is proportional to the magnitude of the source charge, $q$, in the background geometry.
Wormholes are known to allow arbitrary fluxes given by some family of homogeneous solutions to the Maxwell equations in their specific non-trivial spacetime; this refers to J. Wheeler's {\it charge-without-charge} realization \cite{wheeler}. 
For instance, in the considered cylindrical geometry we could consider electric potential fields proportional to the logarithmic functions $\ln{r/r_0}$. These arbitrary solutions come themselves with an associated flux which, in this cylindrically symmetric case, is infinite. 
On the contrary, the flux $\mathcal{F}_{wh}$ we found here is finite and it is not arbitrary, it is given by the presence of a source and the particular topology of the background manifold.

The flux $\mathcal{F}_{wh} = 4\pi q/2$ can be probed as a monopolar charge from the sourceless side of the wormhole; this perspective partially resembles Wheeler's idea of {\it charge-without-charge} -despite it is originated by the source particle.
Some interesting features must be mentioned to complete our discussion. If we put a neutral charge matter content in Maxwell equations, for example, a charge anti-charge pair, then the electric potential in our wormhole spacetime is a superposition of solutions as the one we found, but now producing a total null flux $\mathcal{F}_{\infty}^{q\bar{q}} =0$ as it should be for charge conservation. Note that, in this case, the imprint at each infinity is always $\mathcal{F}^{q\bar{q}}_{\infty+} = \mathcal{F}^{q\bar{q}}_{\infty-} = 0$, irrespective of the positions of the charged particle and the anti-particle. 
Thereby, two configurations can be considered: (i) both particles in the same side of the wormhole producing a flux through the throat $\mathcal{F}_{wh}^{q\bar{q}} =0$, or (ii) the pair is separated so that each particle is driven to a different side of the wormhole, producing $\mathcal{F}_{wh}^{q\bar{q}} = 4 \pi q$ pointing from the positive charge side to the negative one. 
In case (ii) the spacetime started without fluxes and ended up with a flux through the throat and with non-null charge in each submanifold. No physical contradiction appears by arranging this configuration; the imprints at each infinity remain both null \cite{whs3}.
In the last scenario the drain hole effect through the throat measures exactly the flux corresponding to the charge magnitud $q$ of the pair, resembling a {\it flux-without-flux} configuration. 
From another perspective, if we lose sight of the presence of the throat and examine only one side of the wormhole where $\mathcal{F}_{\infty+}^{q\bar{q}} = 0$ (or $\mathcal{F}_{\infty-}^{q\bar{q}} = 0$), we find a {\it charge-without-flux} picture;
the particle interacts electrically over its submanifold, but no asymptotic flux is measured.

\end{document}